\documentclass[aps,pra,twocolumn,superscriptaddress]{revtex4-1}

\usepackage{graphicx}
\usepackage{booktabs}
\AtBeginDocument{
	\heavyrulewidth=.08em
	\lightrulewidth=.05em
	\cmidrulewidth=.03em
	\belowrulesep=.65ex
	\belowbottomsep=0pt
	\aboverulesep=.4ex
	\abovetopsep=0pt
	\cmidrulesep=\doublerulesep
	\cmidrulekern=.5em
	\defaultaddspace=0.2em
}

\usepackage{longtable}
\usepackage{threeparttable,array,booktabs,calc}
\usepackage{graphicx}
\usepackage{dcolumn}
\usepackage{bm}

\usepackage[version=4]{mhchem}

\usepackage{hyperref}

\usepackage{epstopdf} 

\begin{document}

\title[He*-Li(S,P) collisions]{Suppression of Penning ionization by orbital angular momentum conservation}

\author{Katrin Dulitz}
\email{katrin.dulitz@physik.uni-freiburg.de.}
\affiliation{Institute of Physics, University of Freiburg, Hermann-Herder-Str. 3, 79104 Freiburg, Germany}
\author{Tobias Sixt}
\affiliation{Institute of Physics, University of Freiburg, Hermann-Herder-Str. 3, 79104 Freiburg, Germany}
\author{Jiwen Guan}
\affiliation{Institute of Physics, University of Freiburg, Hermann-Herder-Str. 3, 79104 Freiburg, Germany}
\author{Jonas Grzesiak}
\affiliation{Institute of Physics, University of Freiburg, Hermann-Herder-Str. 3, 79104 Freiburg, Germany}
\author{Markus Debatin}
\affiliation{Institute of Physics, University of Freiburg, Hermann-Herder-Str. 3, 79104 Freiburg, Germany}
\author{Frank Stienkemeier}
\affiliation{Institute of Physics, University of Freiburg, Hermann-Herder-Str. 3, 79104 Freiburg, Germany}

\date{\today}

\begin{abstract}	
The efficient suppression of autoionizing collisions is a stringent requirement to achieve quantum degeneracy in metastable rare gases. Here, we report on the suppression of Penning ionization in collisions between metastable He and laser-excited Li atoms. Our results show that the suppression is achieved by the conservation of both the total electron spin and $\Lambda$, i.e., the projection of the total molecular orbital angular momentum along the internuclear axis. Our findings suggest that $\Lambda$ conservation can be used as a more general means of reaction control, in particular, to improve schemes for the simultaneous laser cooling and trapping of metastable He and alkali atoms. 
\end{abstract}

\pacs{}
\keywords{}

\maketitle

\section{Introduction}
Since atoms in electronically excited, long-lived (``metastable'') states have various applications in cold chemistry, atomic optics, statistical physics and in surface science, the study of autoionization dynamics in metastable rare gas collisions has a long research tradition \cite{Siska1993, Vassen2012, Onellion1984, Harada1997}.

Reactive scattering between a metastable atom A$^*$ and a collision partner B, whose ionization potential (IP) is lower than the energy of A$^*$, can lead to the ionization of B (Penning ionization, PI) or to the formation of a molecular ion AB$^+$ and an electron (associative ionization, AI):
\begin{equation}
\mathrm{A^*+B \rightarrow [AB]^*\rightarrow}
\left\{  
\begin{matrix}
\mathrm{A + B^+ + e^-}\\
\mathrm{AB^+ + e^-}
\end{matrix}
\right.
\end{equation}

Recently, it has been shown that the relative rates of PI and AI can be influenced by changing the relative orientation of the interacting atomic orbitals \cite{Gordon2017, Zou2018, Gordon2018}. Autoionization reactions can also be studied and controlled by other means, e.g., by preparing the colliding species in a specific internal state, by lowering the collision energy, as well as by implementing coherent control schemes \cite{Arango2006, Omiste2018}. For instance, in sub-Kelvin collisions of metastable rare gas species, where only few partial waves are involved, the low collision energy has enabled the observation of orbiting resonances \cite{Henson2012, Jankunas2015b}. Such experiments provide opportunities to understand the nature of the intermediate collision complex, which is not possible in conventional scattering experiments.

To prevent rapid trap losses, reaction control is particularly important to achieve quantum degeneracy in an ultracold gas of metastable atoms. For instance, autoionizing collisions can be avoided by preparing the atoms in spin-stretched magnetic substates \cite{Vassen2012}. This preparation scheme exploits the fact that, according to Wigner's spin-conservation rule \cite{Wigner1928}, the total electron spin is a conserved quantity to a good approximation. For example, autoionizing collisions are strongly suppressed inside a spin-polarized sample of He($2^3\mathrm{S}_1$) \cite{Herschbach2000}. In this case, collisions can only happen within the quasimolecular symmetry $^5\Sigma^+_\mathrm{g}$ with a total spin of 2. This state cannot autoionize, since the total spin of the products cannot be greater than 1. As a result, the autoionization rate of spin-polarized atoms is about five orders of magnitude lower compared to the unpolarized case \cite{Shlyapnikov1994}. For He($2^3\mathrm{S}_1$)-Rb($5^2\mathrm{S}_{1/2}$) collisions, electron-spin statistical effects were also found to dominate the reactivity \cite{Byron2010, Flores2016}.

It is known that, for single-species collisions at ultracold temperatures, the electronic excitation of one collision partner can lead to an enhancement or to a suppression of the ionization rate depending on the detuning of the excitation light with respect to the atomic resonance frequency \cite{Weiner1999}. The rate enhancement for red-detuned light can be rationalized by a more attractive long-range potential $V(R)$, which changes from $\propto 1/R^6$ to $\propto 1/R^3$ upon electronic excitation \cite{Mastwijk1998,Mastwijk1998a}. For excitation with blue-detuned light, the atoms are brought to a repulsive state, which prevents the atoms from approaching one another at close distance \cite{Walhout1995, Zilio1996}. For two-species collisions, such as those presented in this article, the autoionization rates are not expected to change upon electronic excitation, since the shape of the attractive long-range potential remains $V(R)\propto 1/R^6$ even when one of the colliding atoms is in an excited state. Nevertheless, for two-species Penning collisions between metastable He atoms (in the $2^1\mathrm{S}_0$ and $2^3\mathrm{S}_1$ states) and Li atoms, we observe a marked decrease in reactivity upon electronic excitation of Li from the $2^2\mathrm{S}_{1/2}$ electronic ground state to the $2^2\mathrm{P}_{1/2,3/2}$ states. 

In this article, we present the results of these experimental studies. We illustrate that our findings can be brought into remarkable agreement with a model which is based on electron-spin conservation and on the conservation of the projection of the orbital angular momentum along the internuclear axis, $\Lambda$. Our research extends previous work by Morgner and co-workers, who have suggested that $\Lambda$ conservation can explain the autoionization rates observed for different metastable-rare-gas collision systems \cite{Hoffmann1979, Lorenzen1983, Siska1993}. While the findings of previous experimental measurements were only indicative, our results provide the first direct experimental evidence of this Penning suppression mechanism. 

\section{Experiment}
\begin{figure}[h!]
	\includegraphics{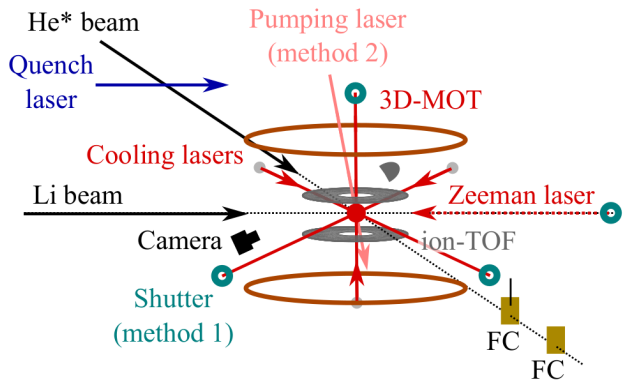}
	\caption{Schematic drawing of the experimental setup (not to scale). FC = Faraday cup, ion-TOF = ion-time-of-flight detector}
	\label{fig:figexpsetup}
\end{figure}

Most parts of the experimental setup have already been described elsewhere \cite{Grzesiak2019, Guan2019}. Therefore, only relevant details are given here. The setup, which is schematically depicted in Fig. \ref{fig:figexpsetup}, consists of a pulsed supersonic beam for $^4$He atoms and a magneto-optical trap (MOT) for ultracold $^7$Li atoms (IP = 5.4 eV [3.5 eV] by ionization from the $2^2\mathrm{S}_{1/2}$ [$2^2\mathrm{P}$] state \cite{NIST_ASD}). An electron-seeded He discharge at the front of a pulsed valve (30 $\mu$s pulse duration, 7 Hz repetition rate) produces a mixture of metastable He atoms (denoted as He$^*$ hereafter) in the long-lived, electronically excited $2^1\mathrm{S}_0$ (20.6 eV \cite{NIST_ASD}) and $2^3\mathrm{S}_1$ (19.8 eV \cite{NIST_ASD}) states by electron impact from the $1^1\mathrm{S}_0$ electronic ground state. The He$^*$ beam intensity and its mean velocity ($v_{\mathrm{He}^*}$ = $1850\,\pm\,20$ m/s) are measured on two gold-coated Faraday cup detectors located downstream from the Li-MOT. In order to determine the role of the He($2^1\mathrm{S}_0$) and He($2^3\mathrm{S}_1$) states on the autoionization rates, we optically quench the atoms in the $2^1\mathrm{S}_0$ state to the electronic ground state using \mbox{46 mW} of diode laser radiation resonant with the $4^1\mathrm{P}_1\,\leftarrow\,2^1\mathrm{S}_0$ transition at 397 nm \cite{Guan2019}.
The quenching efficiency is \mbox{$\geq$ 99 \%}. The He$^*$ singlet-to-triplet ratio is determined using Faraday cup detection taking into account the secondary electron ejection efficiencies of He($2^1\mathrm{S}_0$) and He($2^3\mathrm{S}_1$) on gold-plated surfaces \cite{Woodard1978}.

Li atoms are continuously laser-cooled via the $2^2\mathrm{P}_{3/2}\,\leftarrow\,2^2\mathrm{S}_{1/2}$ transition (D$_2$ line) at 671 nm in a setup which consists of a Li oven, a Zeeman slower for Li deceleration and a 3D-MOT. Optical cycling is achieved by continuously exciting the atoms from both hyperfine states of the ground state ($F=1,\,2$) into the $2^2\mathrm{P}_{3/2}$ excited state. Since the hyperfine splitting in the $2^2\mathrm{P}_{3/2}$ state (hyperfine coupling constant $A \approx 3$ MHz \cite{Orth1975}) is of similar magnitude as the natural linewidth of the D$_2$ transition ($\gamma$ = $2\pi\,\cdot\,5.872$ MHz \cite{McAlexander1996}), only two distinct laser frequencies are required to achieve optical cycling. For optical excitation from the $F=2$ and $F=1$ states, laser intensities $I_0 \approx$ 67 mW/cm$^2$ and 49 mW/cm$^2$ are used, respectively.

We use two experimental techniques to distinguish between Li-ground-state and Li-excited-state collisions with He$^*$. In the first scheme (denoted as method 1), mechanical shutters (SRS, SR475, 40 $\mu$s closing time) are used to block the light for the MOT lasers and for the Zeeman slower in front of the vacuum chamber, respectively. The shutters are closed just before \mbox{($\leq$ 150 $\mu$s)} and after the He$^*$ atoms have collided with the Li atoms (see Fig. \ref{fig:fig1}), respectively. Under these conditions, the spatial expansion of the Li cloud is negligible (cf. Ref. \cite{Grzesiak2019}). In between different measurement series, the population in the Li($2^2\mathrm{P}_{3/2}$) state is varied by adjusting the detuning for the MOT lasers (-12 MHz $\geq \delta \geq$ -50 MHz). The Zeeman laser light is set to a large detuning ($\delta$ = -60 MHz) and thus only contributes marginally to the Li-excited-state population.

The second scheme (denoted as method 2) relies on the use of a separate laser system for the optical excitation of Li atoms via the $2^2\mathrm{P}_{3/2}\,\leftarrow\,2^2\mathrm{S}_{1/2}$ transition (D$_2$ line) or via the $2^2\mathrm{P}_{1/2}\,\leftarrow\,2^2\mathrm{S}_{1/2}$ transition (D$_1$ line), respectively. Here, the MOT and Zeeman lasers are continuously operated at detunings of $\delta$ = -50 MHz and -60 MHz, respectively, which results in a small fraction of Li atoms in the $2^2\mathrm{P}_{3/2}$ state ($\approx$ 4 \%). Two laser frequencies are used for the resonant optical excitation from the $2^2\mathrm{S}_{1/2}$($F=1,\,2$) states.
Optical excitation via the D$_2$ line is achieved via the same transitions which are also used for the laser cooling and trapping of Li. For excitation via the D$_1$ line ($A \approx 46$ MHz for the $2^2\mathrm{P}_{1/2}$ state \cite{Orth1975}), the laser frequencies are tuned such that the $2^2\mathrm{P}_{1/2}$($F'=2$) state is excited. To prevent a depletion of the MOT by radiation pressure, the laser light for optical excitation ($I_0 \approx$ 250 mW/cm$^2$ for each laser frequency) is admitted into the He$^*$-Li interaction zone for a duration of 5 $\mu$s (cf. Fig. \ref{fig:fig2}) using acousto-optical modulators ($\approx\,$0.5 $\mu$s rise time). During this time interval, the position of the trapped cloud of Li atoms, which is monitored using fluorescence imaging with a CCD camera, does not change.

Reactive collisions between He$^*$ and Li atoms are studied in the center of the MOT. This is achieved by continuously extracting all ions, which are produced during the He$^*$-Li interaction time, onto a channel electron multiplier. The detector is operated in ion counting mode and the events are assigned to discrete time intervals (2 $\mu$s-long time bins for method 1, 500 ns-long time bins for method 2).
After each reaction rate measurement, an ion trace without Li is taken to allow for background subtraction of all ions not produced by He$^*$-Li collisions.
The AI/PI ratio (obtained using ion time-of-flight detection) is typically \mbox{$\leq\,2\,$\%}, so that our measurements are mostly sensitive to PI. Relative rate measurements are carried out by alternating the quench laser on and off on a shot-by-shot basis. For method 1, the falling edge of the shutters is delayed by $\Delta t$ = 150 $\mu$s for each third and fourth shot of the pulsed valve. For method 2, the switching of the acousto-optical modulators is timed such that the optical excitation of Li only occurs at each third and fourth valve shot.

\section{Results and Discussion}
The He$^*$-Li reaction kinetics can be described as follows:
\begin{equation}
\frac{\mathrm{d}[\mathrm{I}^+]_i}{\mathrm{d}t} = k_{i}[\mathrm{He}^*]_{i,t}[\mathrm{Li}]_{i,0},
\label{eq:rateeq}
\end{equation}
where $k_{i}$ denotes the rate coefficient for a specific electronic-state combination $i$, $[\mathrm{I}^+]_i$ is the state-dependent density of the product ions (Li$^+$ and HeLi$^+$) and $[\mathrm{He}^*]_{i,t}$ is the state- and time-dependent density of He$^*$. The Li atomic density $[\mathrm{Li}]_{i,0}$, related to a specific electronic-state combination $i$, is assumed to be constant. The determination of absolute He$^*$ and Li number densities is very difficult and prone to large systematic uncertainties. For this reason, we have determined rate coefficient \textit{ratios} for different He$^*$-Li electronic-state combinations.

\begin{figure}[h!]
	\includegraphics{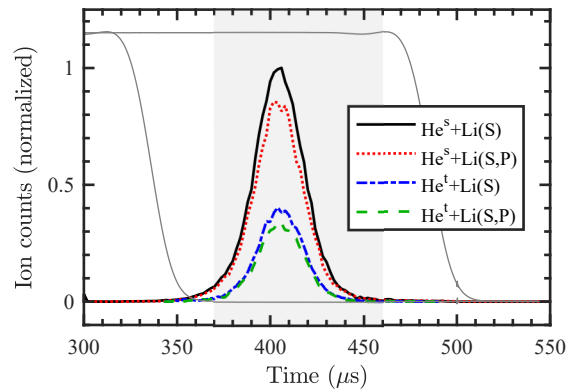}
	\caption{Measured ion yields for electronic-state-selected He$^*$-Li collisions (obtained using method 1) at $\delta$ = -12 MHz as a function of time delay with respect to the valve trigger (thick lines). In the legend, He$^\mathrm{s}$ and He$^\mathrm{t}$ are abbreviations for the He($2^1\mathrm{S}_0$) and He($2^3\mathrm{S}_1$) states, respectively. Li(S) and Li(S, P) denote Li($2^2\mathrm{S}_{1/2}$) and a mixture of Li($2^2\mathrm{S}_{1/2}$) and Li($2^2\mathrm{P}_{3/2}$), respectively. The thick, dash-dotted blue and dashed green curves are scaled with respect to the singlet-to-triplet ratio in the He$^*$ beam to aid the comparison between relative signal intensities. The integration time window used for the calculation of rate coefficients is indicated by the gray shading. Traces of MOT laser stray light (thin gray lines), recorded using a fast photodiode, illustrate the shutter closing characteristics for the two shutter time delays at which the traces were taken.}
	\label{fig:fig1}
\end{figure}

Fig. \ref{fig:fig1} shows measured time traces of the total ion yield for electronic-state-selected He$^*$-Li collisions at a laser detuning of $\delta$ = -12 MHz obtained using method 1. Under these conditions, around 1/3 of the Li atoms are excited to the $2^2\mathrm{P}_{3/2}$ state. The overall time dependence of the ion signal intensity reflects the He$^*$ flux from the pulsed valve (cf. Eq. \ref{eq:rateeq}). This is confirmed by comparing the ion traces with the signal intensities on the Faraday cup detectors (not shown).
A comparison of the relative ion signal intensities in \mbox{Fig. \ref{fig:fig1}} shows that the reactivity of He($2^3\mathrm{S}_1$) is lower than the reactivity of He($2^1\mathrm{S}_0$). Likewise, the reactivity of Li($2^2\mathrm{P}_{3/2}$) is found to be lower than that of Li($2^2\mathrm{S}_{1/2}$).

To extract the rate coefficient ratios, the measured ion signals are time integrated and normalized to the He($2^1\mathrm{S}_0$) and He($2^3\mathrm{S}_1$) fluxes obtained from the respective Faraday cup traces. Relative steady-state populations of Li($2^2\mathrm{P}_{3/2}$) with respect to Li($2^2\mathrm{S}_{1/2}$) are obtained from a rate model of the optical excitation process in Li, taking into account all electric-dipole-allowed transitions between the hyperfine states, the natural linewidth, the experimentally determined laser intensities and the influence of the MOT magnetic field on the transition frequencies.

\begin{figure}[h!]
	\includegraphics{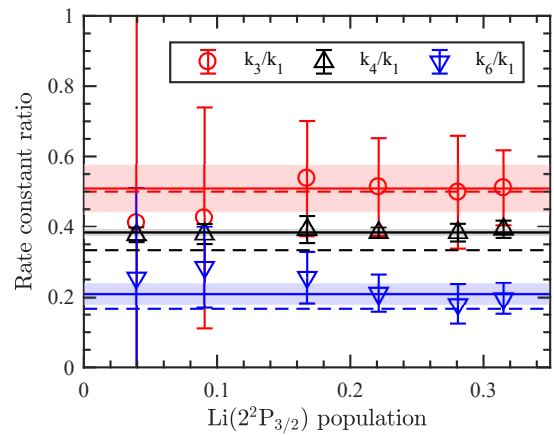}
	\caption{Experimentally measured rate coefficient ratios $k^\mathrm{exp}_{i}/k^\mathrm{exp}_{1}$ (where $i$ = 3, 4, 6; see Tab. \ref{tab:reactivities} for notation), obtained using method 1, as a function of Li($2^2\mathrm{P}_{3/2}$) population relative to Li($2^2\mathrm{S}_{1/2}$) (markers). The solid lines and the shadings are weighted means of the rate coefficient ratios and the resultant standard deviations, respectively. The uncertainties are statistical only ($2\sigma$). The ratios $k_{i,S,\Lambda}/k_{1,S,\Lambda}$, which are calculated from the ratio of autoionizing states to the total number of states assuming electron-spin and $\Lambda$ conservation, are shown as dashed lines for comparison.}
	\label{fig:fig1_supp}
\end{figure}

We have checked the consistency of our results, obtained for method 1, by taking data at different MOT laser detunings corresponding to different Li($2^2\mathrm{P}_{3/2}$) populations. Fig. \ref{fig:fig1_supp} illustrates that the resultant rate coefficient ratios are independent of the Li($2^2\mathrm{P}_{3/2}$) population. The fluctuation of $k^\mathrm{exp}_{3}/k^\mathrm{exp}_{1}$ and $k^\mathrm{exp}_{6}/k^\mathrm{exp}_{1}$ (see Tab. \ref{tab:reactivities} for notation) at small Li($2^2\mathrm{P}_{3/2}$) populations may be attributed to the low signal-to-noise ratio associated with these specific rate determinations.

\begin{figure}[h!]
	\includegraphics{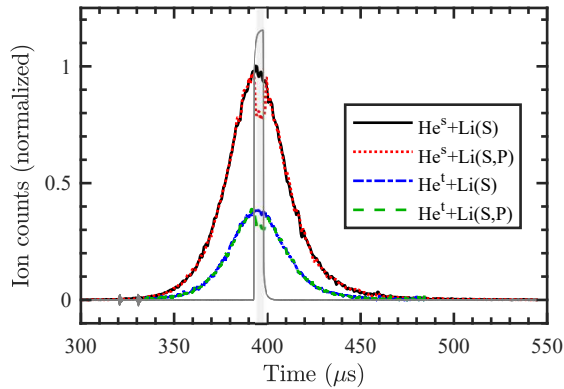}
	\caption{Measured ion yields for state-selected He$^*$-Li collisions as a function of time delay with respect to the valve trigger (thick lines). In contrast to Fig. \ref{fig:fig1}, Li was optically excited to the $2^2\mathrm{P}_{1/2}$ state using method 2. Therefore, the label ``Li(S, P)" in the legend denotes a mixture of Li($2^2\mathrm{S}_{1/2}$) and Li($2^2\mathrm{P}_{1/2}$). The timing characteristics of the laser light for Li optical excitation, measured using a fast photodiode, are illustrated by a thin gray line.}
	\label{fig:fig2}
\end{figure}

Using method 2, we can probe the change in autoionization rates upon laser excitation of Li to the $2^2\mathrm{P}_{3/2}$ or to the $2^2\mathrm{P}_{1/2}$ state, respectively. In both cases, we observe decreased ion signal intensities as soon as the Li population in the excited state is increased. The measured ion traces for He*-Li($2^2\mathrm{P}_{1/2}$) collisions are presented in Fig. \ref{fig:fig2} for comparison.

All experimentally determined He$^*$-Li rate coefficient ratios are summarized in Tab. \ref{tab:reactivities}. The given overall uncertainties include both statistical and systematic errors, such as uncertainties in the determination of the Li-excited-state population and the singlet-to-triplet ratio. Owing to the short additional excitation pulse for method 2, the population dynamics of the optical excitation process are more involved  than for method 1. Therefore, we only assume upper limits for the corresponding Li-excited-state populations and for the rate coefficient ratios in Tab. \ref{tab:reactivities}, respectively.
\begin{table*}[hbt!]
	\caption{\label{tab:reactivities}Summary of molecular-symmetry-based and experimentally determined He$^*$-Li rate coefficient ratios for different electronic-state combinations (numbered as $i\,=\,1-6$); see main text for a detailed assignment of the quasi-molecular states. The ratios $k_{i,S}/k_{1,S}$ ($k_{i,S,\Lambda}/k_{1,S,\Lambda}$) are calculated from the number of autoionizing states with respect to the total number of states assuming electron-spin conservation (electron-spin and $\Lambda$ conservation). The experimentally determined rate coefficient ratios $k^\mathrm{exp}_{i}/k^\mathrm{exp}_{1}$ (including overall uncertainties), obtained using methods 1 and 2, are presented in the last three columns.}
	\begin{tabular}{|r|l|l|c|c|c|c|c|}
		\toprule
		& & & & & \multicolumn{3}{c|}{$k^\mathrm{exp}_{i}/k^\mathrm{exp}_{1}$}\\
		$i$ & Asymptotic states & Quasi-molecular states & $k_{i,S}/k_{1,S}$ & $k_{i,S,\Lambda}/k_{1,S,\Lambda}$ & Method 1 & \multicolumn{2}{c|}{Method 2}\\
		    &                   &                        &                   &                                   &   & via Li D$_2$ line& via Li D$_1$ line\\
		\midrule
		1 & He($2^1\mathrm{S}_0$)+Li($2^2\mathrm{S}_{1/2}$) & $^2\Sigma^+$                                & 1   & 1   & 1 & 1 & 1\\
		\rule{0pt}{10pt}
		2 & He($2^1\mathrm{S}_0$)+Li($2^2\mathrm{P}_{1/2}$) & $^2\Pi_{1/2}$                               & 1   & 0   & -- & -- & $\leq$ 0.41\\
		\rule{0pt}{10pt}
		3 & He($2^1\mathrm{S}_0$)+Li($2^2\mathrm{P}_{3/2}$) & $^2\Sigma^+$, $^2\Pi_{3/2}$                 & 1   & 1/2 & $0.51_{-0.07}^{+0.07}$ & $\leq$ 0.44 &--\\
		\rule{0pt}{10pt}
		4 & He($2^3\mathrm{S}_1$)+Li($2^2\mathrm{S}_{1/2}$) & $^{2}\Sigma^+$, $^{4}\Sigma^+$              & 1/3 & 1/3 & $0.38_{-0.02}^{+0.06}$ & $0.39_{-0.04}^{+0.07}$ & $0.42_{-0.04}^{+0.07}$\\
		\rule{0pt}{10pt}		
		5 & He($2^3\mathrm{S}_1$)+Li($2^2\mathrm{P}_{1/2}$) & $^{2}\Pi_{1/2}$, $^{4}\Pi_{1/2,3/2}$        & 1/3 & 0   & -- & -- & $\leq$ 0.22\\
		\rule{0pt}{10pt}	
		6 & He($2^3\mathrm{S}_1$)+Li($2^2\mathrm{P}_{3/2}$) & $^{2}\Pi_{3/2}$, $^{4}\Pi_{1/2,5/2}$, $^{2}\Sigma^+$, $^{4}\Sigma^+$
		& 1/3 & 1/6 & $0.21_{-0.03}^{+0.04}$ & $\leq$ 0.18 & --\\
		\bottomrule
	\end{tabular}
\end{table*}

\begin{figure}[h!]
	\includegraphics{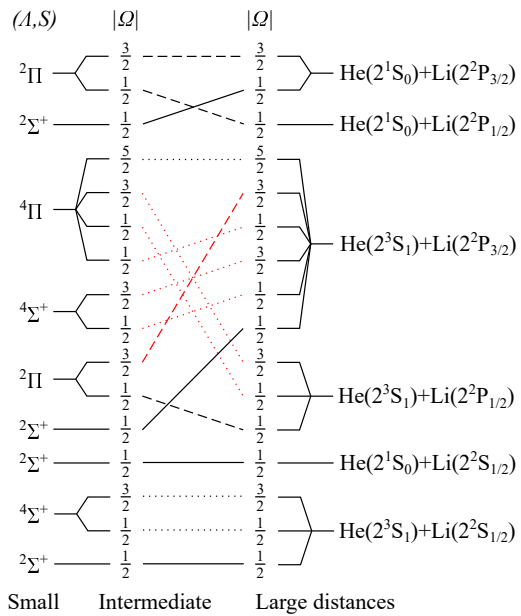}
	\caption{Proposed coupling scheme for He($2^3\mathrm{S}_{1}$,$2^1\mathrm{S}_{0}$)+Li($2^2\mathrm{S}_{1/2}$,$2^2\mathrm{P}_{1/2,3/2}$). Correlations which lead to quartet states of the molecule are indicated as dotted lines. The dashed and solid lines represent correlations to molecular states of $^2\Pi$ and $^2\Sigma^{+}$ symmetry, respectively. In our model, only states of $^2\Sigma^{+}$ symmetry lead to Penning ionization. Correlations shown in red color are tentative only.}
	\label{fig:correlationdiag}
\end{figure}

The correlation diagram in Fig. \ref{fig:correlationdiag} illustrates the assignment of molecular states given in Tab. \ref{tab:reactivities}. The energetic ordering was conceived from the potential energy curves for the He-Li system in Refs. \cite{Ruf1987, Kimura1990, Movre2000} and from the energies of the spin-orbit states in Li. The assignment of the absolute value of the total angular momentum quantum number $|\Omega|$ is based on energy considerations and on our experimental observations.

To understand the experimental results, several rate-influencing effects have to be considered. This includes spin-statistical effects, atomic orbital overlap and -- since autoionization can be regarded as a Franck-Condon-type process -- the shape of the potential curves for the entrance and the exit channels.

Since the exit channel has a total electron spin of $S\,=\,1/2$, only channels with a total electron spin $S\,=\,1/2$ can autoionize owing to electron spin conservation (Wigner's spin conservation rule). The collision complex for He($2^1\mathrm{S}_0$)+Li($2^2\mathrm{S}_{1/2}$) collisions is of $^2\Sigma^{+}$ symmetry and thus fulfills this requirement. In He($2^3\mathrm{S}_1$)+Li($2^2\mathrm{S}_{1/2}$) collisions, two states of $^2\Sigma^{+}$ symmetry and four states of $^4\Sigma$ symmetry are formed. This implies that all (only 1/3) of the He atoms in the $2^1\mathrm{S}_0$ state ($2^3\mathrm{S}_1$ state) can autoionize. The rate coefficient ratio $k_{4}/k_{1}$ = 0.26$\,\pm\,$0.08, previously determined by Ruf et. al. \cite{Ruf1987}, is (within the error) in agreement with this spin-statistical argument. Our measured values for $k^\mathrm{exp}_{4}/k^\mathrm{exp}_{1}$ = $0.38_{-0.02}^{+0.06}$ and $0.39_{-0.04}^{+0.07}$ (see Tab. \ref{tab:reactivities}) are higher than the value of 1/3 expected from spin statistics. This indicates that, to a small extent, the observed rate coefficient ratios are influenced by other processes, such as a different shape of the short-range potentials \cite{Movre2000}.

Our rate coefficient ratios for He$^*$-Li($2^2\mathrm{P}_{3/2}$) collisions ($k^\mathrm{exp}_{3,6}/k^\mathrm{exp}_{1}$) are a factor of two lower than what would be expected from electron-spin conservation only. In He($2^1\mathrm{S}_0$)-Li($2^2\mathrm{S}_{1/2}$) and He($2^1\mathrm{S}_0$)-Li($2^2\mathrm{P}_{1/2}$) collisions, only molecular states of doublet symmetry are involved. Therefore, the rate coefficients for both processes should be the same, i.e. $k_{1,S} = k_{2,S}$. However, from Fig. \ref{fig:fig2} and Tab. \ref{tab:reactivities}, it is obvious that $k^\mathrm{exp}_{1} > k^\mathrm{exp}_{2}$. 
Our results can be explained if we assume that both the electron spin and the projection of the total molecular orbital angular momentum along the internuclear axis, $\Lambda$, are conserved in the collision process. $\Lambda$ conservation implies that only quasi-molecular states of $\Sigma$ symmetry can autoionize, i.e., states of $\Pi$ symmetry do not lead to PI and AI. In Tab. \ref{tab:reactivities}, we present estimates of the He$^*$-Li rate coefficient ratios including both electron-spin and $\Lambda$ conservation. For this estimate, we assume that all states of $^2\Sigma^{+}$ ($^{2,4}\Pi$, $^4\Sigma^{+}$) symmetry autoionize (do not autoionize). The thus obtained rate coefficient ratios $k_{i,S,\Lambda}/k_{1,S,\Lambda}$ are in striking agreement with the experimental results obtained from method 1. As mentioned above, the Li-excited-state populations could not be determined for method 2. Therefore, these latter results are qualitative only. However, the marked decrease of the ion yield upon Li excitation (see Fig. \ref{fig:fig2}) is clearly in line with $\Lambda$ conservation.

Some correlations in Fig. \ref{fig:correlationdiag} (shown in red color) are tentative only. In our model calculations including electron spin and $\Lambda$ conservation, we assume that all states with $^2\Pi$, $^4\Pi$ and $^4\Sigma$ symmetry do not autoionize. Therefore, an incorrect assignment of these states does not affect the rate coefficient ratios $k_{1-6,S,\Lambda}/k_{1,S,\Lambda}$ given in Tab. \ref{tab:reactivities}. Considering electron spin conservation only, an incorrect assignment would leave $k_{1-4,S}/k_{1,S}$ unchanged, while $k_{5,S}/k_{1,S}$ = 2/3 and $k_{6,S}/k_{1,S}$ = 1/5 would also be possible. These rate coefficient ratios are also not consistent with our experimental values. Therefore, our results cannot be explained using electron spin conservation only, even in case of an erroneous assignment of some molecular states.

$\Lambda$ conservation can be rationalized in terms of orbital overlap. Since autoionization processes involving He$^*$ predominantly occur via electron exchange \cite{Siska1993}, the reaction rate is high when there is constructive overlap between the 1s core orbital of the He$^*$ atom and the Li valence shell atomic orbital. If the Li atom is excited from the 2s orbital to a 2p$_{x,y}$ orbital, the overlap integral -- and thus the ionization rate -- is small.

The underlying reason for $\Lambda$ conservation in He$^*$-Li collisions is the small spin-orbit coupling in the entrance and exit channels. In Li, the energy splitting between the two spin-orbit states is \mbox{$\Delta E$($2^2\mathrm{P}_{1/2}$-$2^2\mathrm{P}_{3/2}$) = 0.34 cm$^{-1}$ \cite{Sansonetti1995}}, which corresponds to a spin-orbit-interaction time which is more than two orders of magnitude higher than the collision time of $\approx$ 250 fs \footnote{The collision time is estimated from a classical trajectory calculation along the He(2$^1\mathrm{S}_0$)-Li(2$^2\mathrm{S}_1/2$) potential energy curve \cite{Movre2000} from large distance until the classical turning point. A collision temperature of 530 K is assumed which corresponds to the conditions used in the experiment.} at thermal energies. This implies that the electron leaves the collision complex long before a significant coupling to the internal degrees of freedom can occur.

Compared to electron-spin and $\Lambda$ conservation, the influence of other rate-influencing effects is expected to be small for several reasons. Firstly, for He$^*$-Li(S,P) collisions, only the X$^1\Sigma$ state of HeLi$^+$ is energetically accessible at the collision energies of our experiment. This state corresponds to the He($1^1\mathrm{S}_{0}$)+Li$^+$($1^1\mathrm{S}_{0}$) asymptote.
The next higher-lying states (correlating with the A$^1\Sigma^+$ and a$^3\Sigma^+$ states) are $\approx$ 19 eV higher in energy \cite{Hiyama1992}. Therefore, the admixture of other electronic states is negligible. Secondly, the hyperfine state of the Li atom should, to a first approximation, not influence the autoionization rate, because the hyperfine interaction in Li is much smaller than the spin-orbit interaction. Moreover, the potential shape has little influence on the relative autoionization rates, since the positions of the He$^*$-Li(S,P) potential minima \cite{Kimura1990} and the long-range potentials \cite{Zhang2012} are similar. This is further supported by the results of classical capture calculations detailed below.

\subsection{Capture rate calculations}
We have calculated the rate coefficients for different He$^*$-Li interaction channels using a classical capture model, in which only long-range interactions are assumed to contribute to chemical reactivity \cite{Chernyi2002}. In this simplified theoretical approach, it is assumed that the reaction probability for a specific channel $N$ is unity (zero) if the energy of the collision pair is above (below) the centrifugal barrier of the effective long-range potential
\begin{equation}
	V_N(R) = \sum_{n}-\frac{C_{n,N}}{R^n} + \frac{(\mu v_{\mathrm{rel}} b)^2}{2\mu R^2}.
	\label{eq:Veff}
\end{equation}
Here, $R$ is the internuclear distance, $C_{n,N}$ (with $n = 6, 8, 10$) are the channel-dependent dispersion coefficients, $\mu$ is the reduced mass, $v_{\mathrm{rel}}$ is the relative velocity and $b$ is the impact parameter.

Capture rate coefficients for each electronic-state combination $i$
\begin{equation}
	k_{i}^\mathrm{c} = \sum_{\mathrm{channels}}\pi b_{N,\mathrm{max}}^2 v_{\mathrm{rel}}
	\label{eq:kic}
\end{equation}
are numerically calculated using the dispersion coefficients $C_{n,N}$ obtained by Zhang et al. \cite{Zhang2012} (see Tab. \ref{tab:Cn}). Here, $b_{N,\mathrm{max}}$ is the impact parameter for a specific reaction channel, for which the barrier maximum is equal to the collision energy.

\begin{table*}[hbt!]
	\caption{\label{tab:Cn} Dispersion coefficients for He$^*$-Li interactions used for the classical-capture calculation of rate coefficient ratios (taken from Zhang et al. \cite{Zhang2012}).}
	\begin{tabular}{|l|l|r|r|r|}
		\toprule
		Asymptotic states & Quasi-molecular states & $C_6$ (in a.u.) & $C_8$ (in a.u.) & $C_{10}$ (in a.u.)\\
		\midrule
		He($2^1\mathrm{S}_0$)-Li($2^2\mathrm{S}_{1/2}$) & $^{2}\Sigma^+$ & $3.504\cdot10^3$ & $2.633\cdot10^5$ & $3.018\cdot10^7$\\
		He($2^1\mathrm{S}_0$)-Li($2^2\mathrm{P}_{1/2, 3/2}$) & $^{2}\Pi_{1/2, 3/2}$ & $2.663\cdot10^3$ & $5.222\cdot10^5$ & $3.298\cdot10^7$\\
		He($2^1\mathrm{S}_0$)-Li($2^2\mathrm{P}_{1/2, 3/2}$) & $^{2}\Sigma^+$ & $2.048\cdot10^3$ & $2.483\cdot10^6$ & $2.600\cdot10^8$\\
		He($2^3\mathrm{S}_1$)-Li($2^2\mathrm{S}_{1/2}$) & $^{2}\Sigma^+$, $^{4}\Sigma^+$ & $2.090\cdot10^3$ & $1.326\cdot10^5$ & $1.280\cdot10^7$\\
		He($2^3\mathrm{S}_1$)-Li($2^2\mathrm{P}_{1/2, 3/2}$) & $^{2}\Pi_{1/2, 3/2}$, $^{4}\Pi_{1/2, 3/2, 5/2}$ & $7.605\cdot10^2$ & $1.571\cdot10^5$ & $1.034\cdot10^7$\\				
		He($2^3\mathrm{S}_1$)-Li($2^2\mathrm{P}_{1/2, 3/2}$) & $^{2}\Sigma^+$, $^{4}\Sigma^+$ & $-2.190\cdot10^3$ & $1.063\cdot10^6$& $1.230\cdot10^8$\\
		\bottomrule
	\end{tabular}
\end{table*}

\begin{figure}[h!]
	\includegraphics{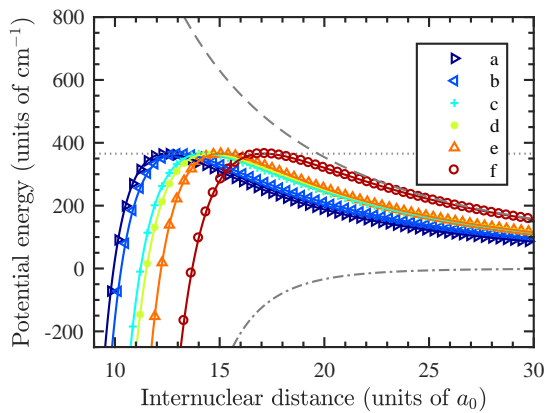}
	\caption{Effective long-range potentials (solid lines with markers) for all He($2^3\mathrm{S}_{1}$,$2^1\mathrm{S}_{0}$)-Li($2^2\mathrm{S}_{1/2}$,$2^2\mathrm{P}_{J}$) channels ($J = 1/2, 3/2$), in which the maximum of the barrier is equal to the collision energy. The legend labels denote the following states: a = He($2^3\mathrm{S}_1$)-Li($2^2\mathrm{P}_{J}$) $^{2,4}\Pi$, b = He($2^3\mathrm{S}_1$)-Li($2^2\mathrm{S}_{1/2}$) $^{2,4}\Sigma$, c = He($2^1\mathrm{S}_0$)-Li($2^2\mathrm{S}_{1/2}$) $^{2}\Sigma$, d = He($2^1\mathrm{S}_0$)-Li($2^2\mathrm{P}_{J}$) $^{2}\Pi$, e = He($2^3\mathrm{S}_1$)-Li($2^2\mathrm{P}_{J}$) $^{2,4}\Sigma$, f = He($2^1\mathrm{S}_0$)-Li($2^2\mathrm{P}_{J}$) $^{2}\Sigma$. In addition to that, the interaction potential $\sum_{n}-C_{n,N}/R^n$ and the centrifugal potential $(\mu v_{\mathrm{rel}} b_{N,\mathrm{max}})^2/(2\mu R^2)$ for system f are also shown as dashed and dash-dotted lines, respectively. The collision energy in the experiment is indicated by a dotted horizontal line.}
	\label{fig:PES}
\end{figure}

The relative velocity and the collision energy are obtained from the experimental conditions of our experiment, where $v_{\mathrm{Li}}\,\approx\,0$, so that $v_{\mathrm{rel}} \approx v_{\mathrm{He}^*}$ = 1850 m/s. Values of $b_{N,\mathrm{max}}$ are obtained numerically by step-wise incrementing the impact parameter until the maximum of the effective potential (Eq. \ref{eq:Veff}) is equal to the collision energy in the experiment. The effective long-range interaction potentials for all He$^*$-Li interaction channels are shown in Fig. \ref{fig:PES}.

The resulting rate coefficient ratios given in Tab. \ref{tab:capturerates} are not in agreement with the experimental data if electron spin conservation is considered only. The results from these calculations can be brought closer into agreement with the experimental results if statistical weights for both electron-spin and $\Lambda$ conservation are included in the capture code. A quantitative agreement is not expected, since classical capture theory can only provide upper bounds for the rate coefficients \cite{Henchman1972}. An accurate quantum-chemical treatment of the He*-Li system, including a calculation of the interaction potentials and the ionization widths for all channels, is required to obtain more detailed information about the underlying reaction dynamics.
\begin{table*}[hbt!]
	\caption{\label{tab:capturerates}Summary of He$^*$-Li rate coefficient ratios obtained from classical-capture calculations. The ratios $k^\mathrm{c}_{i}/ k^\mathrm{c}_{1}$ are calculated using Eq. \ref{eq:kic}. For $k^\mathrm{c}_{i,S}/ k^\mathrm{c}_{1,S}$ ($k^\mathrm{c}_{i,S,\Lambda}/ k^\mathrm{c}_{1,S,\Lambda}$), also electron-spin (electron-spin and $\Lambda$ conservation) is considered.}
	\begin{tabular}{|r|l|l|c|c|c|}
		\toprule
		$i$ & Asymptotic states & Quasi-molecular states & $k^\mathrm{c}_{i}/k^\mathrm{c}_{1}$ & $k^\mathrm{c}_{i,S}/ k^\mathrm{c}_{1,S}$ & $k^\mathrm{c}_{i,S,\Lambda}/k^\mathrm{c}_{1,S,\Lambda}$\\
		\midrule
		1 & He($2^1\mathrm{S}_0$)+Li($2^2\mathrm{S}_{1/2}$) & $^2\Sigma^+$                                & 1    & 1 & 1\\
		\rule{0pt}{10pt}
		2 & He($2^1\mathrm{S}_0$)+Li($2^2\mathrm{P}_{1/2}$) & $^2\Pi_{1/2}$                               & 1.02 & 1.02 & 0\\
		\rule{0pt}{10pt}
		3 & He($2^1\mathrm{S}_0$)+Li($2^2\mathrm{P}_{3/2}$) & $^2\Sigma^+$, $^2\Pi_{3/2}$                 & 1.20 & 1.20 & 0.68\\
		\rule{0pt}{10pt}
		4 & He($2^3\mathrm{S}_1$)+Li($2^2\mathrm{S}_{1/2}$) & $^{2}\Sigma^+$, $^{4}\Sigma^+$              & 0.84 & 0.28 & 0.28\\
		\rule{0pt}{10pt}		
		5 & He($2^3\mathrm{S}_1$)+Li($2^2\mathrm{P}_{1/2}$) & $^{2}\Pi_{1/2}$, $^{4}\Pi_{1/2,3/2}$        & 0.74 & 0.25 & 0\\
		\rule{0pt}{10pt}
		6 & He($2^3\mathrm{S}_1$)+Li($2^2\mathrm{P}_{3/2}$) & $^{2}\Pi_{3/2}$, $^{4}\Pi_{1/2,5/2}$, $^{2}\Sigma^+$, $^{4}\Sigma^+$ & 0.87 & 0.29 & 0.17\\
		\bottomrule
	\end{tabular}
\end{table*}

\section{Conclusion}
We have observed a significant decrease of the He$^*$-Li reaction rate upon electronic excitation of Li from the $2^2\mathrm{S}_{1/2}$ ground state to the $2^2\mathrm{P}_{1/2, 3/2}$ states. Using simple theoretical models based on spin statistical weights and the He$^*$-Li long-range interaction potentials, we have shown that both the electron spin and $\Lambda$ are nearly conserved in He$^*$-Li autoionizing collisions.

We are confident that the reported findings can be applied to a number of autoionizing systems, in which the spin-orbit interaction in the entrance and exit channels is small. This includes reactive collisions between metastable He atoms and electronically excited first-column atoms (H, Li, Na, K \footnote{In He$^*$-K collisions, the autoionization rates are also influenced by resonances with core-excited states in K \cite{Johnson1978}.}) as well as reactions of other metastable atoms (N($2^2\mathrm{P}_{1/2,3/2}$), O($2^1\mathrm{S}_0$, $2^5\mathrm{S}_2$), C($2^5\mathrm{S}_2$)) with light alkali atoms in the P states.

Since the \textit{ab-initio} calculation of ionization widths requires the intricate evaluation of two-center two-electron integrals \cite{OpdeBeek1997}, the approach shown above provides a simple alternative means to interpret the fate of autoionizing collisions, including the assignment of quasi-molecular states. The observed deviations between our measured rate coefficient ratios and the predictions may trigger additional quantum-chemical calculations, and provide further insight into processes not considered here, such as non-Born Oppenheimer effects. Since a total of only five electrons is involved, the He-Li system is particularly suited for accurate quantum-chemical calculations.

Autoionizing collisions involving metastable atoms are also of particular relevance for laser cooling and trapping applications \cite{Vassen2012}. However, the influence of $\Lambda$ conservation has not been considered for these applications yet. For example, in order to reduce trap loss by autoionization in a two-species He$^*$-Li MOT, Li could be laser cooled via the D$_1$ line using a gray molasses scheme \cite{Grier2013}.


\begin{acknowledgments}
Financial support by the German Research Foundation (projects DU1804/1-1 and GRK 2079) and by the Chemical Industry Fund (Liebig Fellowship to K. Dulitz) is gratefully acknowledged. J. Grzesiak is thankful for a scholarship by the International Graduate Academy of the Freiburg Research Services. This work has greatly benefited from discussions with P. {\.Z}uchowski (Toru\'{n}), R. Krems (Vancouver) and M. Mudrich (Aarhus).	
\end{acknowledgments}

%
	
\end{document}